\documentclass[aps,showpacs,pra]{revtex4}
\usepackage{epsfig,amssymb,amscd}
\begin{document}
\bibliographystyle{prsty}

\title{\textbf{Generation of entangled photons by trapped ions in
microcavities under a magnetic field gradient}}
\author{M. Feng $^{1,2}$, Z.J. Deng $^{1,2,3}$, K.L. Gao $^{1,2}$}
\affiliation{$^{1}$ State Key Laboratory of Magnetic Resonance and Atomic and Molecular
Physics, Wuhan Institute of Physics and Mathematics, Chinese Academy of
Sciences, Wuhan, 430071, China \\
$^{2}$ Center for Cold Atom Physics, Chinese Academy of Sciences, Wuhan
430071, China \\
$^{3}$ Graduate School of the Chinese Academy of Sciences, Beijing 100049,
China}
\date{\today}

\begin{abstract}
We propose a potential scheme to generate entangled photons by manipulating
trapped ions embedded in two-mode microcavities, respectively, assisted by a
magnetic field gradient. By means of the spin-spin coupling due to the
magnetic field gradient and the Coulomb repulsion between the ions, we show
how to efficiently generate entangled photons by detecting the internal
states of the trapped ions. We emphasize that our scheme is advantageous to
create complete sets of entangled multi-photon states. The requirement and
the experimental feasibility of our proposal are discussed in detail.
\end{abstract}

\maketitle

\vskip 0.1cm

\section{introduction}

Entangled photons are important sources in quantum communication and quantum
cryptography. To generate entangled photons, we can make use of atomic
cascade decay \cite{aspect}, parametric down conversion in nonlinear
crystals \cite{kwiat}, and exciton emission in a semiconductor quantum dots 
\cite{bensen,loss}. Besides, entangled photons can also be produced by using
polarizing beam splitters \cite{shih}. On the other hand, there have been a
lot of proposals to entangle atoms by detecting the emitted photons, in
which the entanglement of the photons produced by beam splitters is
projected to internal degrees of freedom of the atoms \cite{papers}. Since
photons are flying qubits, this is a way towards future quantum network
based on local atomic qubits.

In this paper, we study a scheme to project the entanglement in atomic
states to the photons emitted from these atoms, a reverse step with respect
to the proposals in \cite{papers}. We noticed a recent work for the same
purpose \cite{beige}, in which the entangled photon pairs can be created
from two distant dipole sources by means of the postselection and
interference effects. In contrast, we create entangled photons by using
modified ion traps, with which the trapped ions interact via spin-spin
coupling produced by a magnetic field gradient and the Coulomb repulsion 
\cite{wunderlich1,wunderlich2,twamley}. The favorable features of our scheme
include: (1) The entangled photons can be generated efficiently, and
complete sets of entangled states of more than two photons are achievable.
Because we put the trapped ions in microcavities with one ion in a
microcavity, by adjusting the cavity decay rate and the detunings (mentioned
specifically below), we can have a high success rate of photon generation.
With more microtrap-microcavity setups added, the entangled states of more photons 
can be obtained straightforwardly. (2) The entanglement of the
trapped ions is made by the spin-spin couplings, i.e., Ising terms, in which
the vibrational modes of the ions remain unchanged throughout the scheme. So
we don't require the ions to be strictly cooled to the vibrational ground
state and our quantum gating for the entanglement generation is in principle
robust to heating of the ions. (3) The collection rate of the emitted
photons from the ions can in principle approach to unity because the trapped
ions are embedded in microcavities, and we collect the leaking photons from
each microcavities. (4) Since the Ising term is widely used for quantum
computing in other systems, e.g., the semiconductor quantum dots \cite{feng}%
, our proposal can in principle be generalized to those systems.

\section{generation of entangled photon pairs}

We first consider the simplest case, i.e., two trapped ions, for example, Yb$%
^{+}$, in a magnetic field gradient embedded respectively in two two-mode
microcavities. These two ions are also confined in two microtraps,
respectively, as discussed in \cite{twamley} and shown in Fig. 1. Although
in principle they are not required to be in the zero point of the trapping
potential, the ions are required to be strictly within Lamb-Dicke limit
regarding laser radiation. If we encode qubits in the levels $|g\rangle$ and 
$|e\rangle$, respectively, according to \cite%
{wunderlich1,wunderlich2,twamley}, the two qubit states will interact via an
Ising coupling, i.e., $H_{I}=-(J/2)\sigma_{z}^{1}\sigma_{z}^{2}$, where $%
\sigma_{z}^{m}=|e\rangle_{m}\langle e| - |g\rangle_{m}\langle g|$ ($m=1,2$
here and hereafter) and 
\begin{equation}
J=\sum_{n=1}^{2} \frac {1}{\nu_{n}} S_{n1}S_{n2} \frac {\partial\omega_{1}}{%
\partial z}\frac {\partial\omega_{2}}{\partial z} (\Delta z_{n})^{2},
\end{equation}
with $S_{nm}$ a unitary transformation matrix related to the Hessian of the
potential, and $\nu_{n}$ and $\Delta z_{n}$ are respectively the frequency
 and the spatial spread of the ground state wavefunction of
the mode $n$ of the ions' collective  vibration. $\omega_{m}=g\mu_{B}B_{m}$
with $B_{m}$ the magnetic field at the position the ion $m$ staying. J can
be changed by adjusting the trap frequencies, the distance between the two
microtraps and the magnetic field gradient \cite{twamley}.

Our main idea is to map the entanglement from the qubit states to the
photons emitted from the ions. We consider the configuration of each ion as
in Fig. 1(a), where for convenience the hyperfine levels are defined to be: $%
(1,1) = |e\rangle$, $(0,0) = |g\rangle$, $(1,0) = |e^{\prime}\rangle$, $%
(1,-1) = |g^{\prime}\rangle$ in $S_{1/2}$, and $(1,0) = |r\rangle$ in $%
P_{1/2}$. $\Omega_{jm}$ and h$_{jm}$ ( here and hereafter j=g,e) are
coupling constants with respect to lasers and cavity modes to the mth ion,
respectively. We assume that the laser radiation and the cavity modes are
detuned from $|r\rangle$ by $\delta_{jm}$. Due to these large detunings, we
have two Raman processes in each cavity with effective Rabi frequencies $%
\tilde{\Omega}_{jm}=\Omega_{jm} h_{jm}/\delta_{jm}$. For simplicity, we
assume $\tilde{\Omega}_{gm} = \tilde{\Omega}_{em} = \tilde{\Omega}_{m}$ by
appropriately adjusting $\Omega_{jm}$, $h_{jm}$ and $\delta_{jm}$.

Considering the cavity decay, we have following non-Hermitian Hamiltonian
for each cavity, 
\begin{equation}
H_{m}= \tilde{\Omega}_{m} ( c_{gm}|g^{\prime}\rangle_{m}\langle g| +
c^{\dagger}_{gm}|g\rangle_{m}\langle g^{\prime}| +
c_{em}|e^{\prime}\rangle_{m}\langle e| +
c^{\dagger}_{em}|e\rangle_{m}\langle e^{\prime}|) - i\kappa_{m}
(c^{\dagger}_{gm}c_{gm} + c^{\dagger}_{em}c_{em}),
\end{equation}
where we suppose the same decay rate $\kappa_{m}$ for the two cavity modes.
Consider the ion to be intially prepared in the state 
\begin{equation}
|\Psi_{0}\rangle_{m} = \frac {1}{\sqrt{2}} ( |g^{\prime}\rangle_{m} +
|e^{\prime}\rangle_{m}) |00\rangle_{m},
\end{equation}
with the vacuum states of $\sigma_{0}$ and $\sigma_{+}$ cavity modes in mth
cavity denoted by $|00\rangle_{m}$. Before any photon leaking out of each
cavity, we can solve Eq. (2) following the idea in \cite{zou}. After an
interaction time $\tau_{m}$ with $\tau_{m}$ satisfying $tan ( \tilde{\Omega}%
_{m}^{\prime}\tau_{m}) = 2\tilde{\Omega}_{m}^{\prime}/\kappa_{m}$ and $%
\tilde{\Omega}_{m}^{\prime}=\sqrt {\tilde\Omega_{m}^{2}-\kappa^{2}_{m}/4}$,
the system evolves to 
\begin{equation}
|\Psi\rangle = \frac {1}{\sqrt{2}} ( |g\rangle_{m}|10\rangle_{m} +
|e\rangle_{m} |01\rangle_{m}),
\end{equation}
with the success probability 
\begin{equation}
P_{m}= \exp {(-\kappa_{m}\tau_{m})}\sin^{2}(\tilde\Omega^{\prime}_{m}\tau_{m})(%
\tilde\Omega_{m}/\tilde\Omega^{\prime}_{m})^{2}.
\end{equation}
When $t > max \{ 1/\kappa_{1}, 1/\kappa_{2}\}$, photons will leak out of the
cavities, which yields the total wavefunction 
\begin{equation}
|\Phi\rangle = \frac {1}{2}\prod_{m=1}^{2}
(|g\rangle_{m}|\sigma_{0}\rangle_{m} + |e\rangle_{m}|\sigma_{+}\rangle_{m}).
\end{equation}
Before we discuss how to collect and store the emitted photons, we first
focus on how to make Bell measurement on the two ions, which projects
entanglement to the two leaking photons no matter where they are. Our steps
of Bell measurement follow the ideas in \cite{riebe,barret,deng}, i.e.,
performing Hadamard and two-qubit gates before detection on the qubit states
by standard fluorescence techniques \cite{blatt}. Due to the Ising coupling
between the ions, we can carry out the two-qubit gate by using the
sophisticated method in NMR quantum computing \cite{deng}, 
\begin{equation}
CNOT_{12}=e^{-i\pi/4}e^{-i(\pi/4)\sigma^{2}_{y}}e^{i(\pi/4)%
\sigma^{1}_{z}}e^{i(\pi/4)\sigma^{2}_{z}}e^{-i(\pi/4)\sigma^{1}_{z}
\sigma^{2}_{z}}e^{i(\pi/4)\sigma^{2}_{y}},
\end{equation}
where the ions 1 and 2 are control and target, respectively, and $%
\sigma_{y(z)}^{m}$ is the Pauli matrix of the qubit states of the ion $m$ as
defined in $H_{I}$. The Hadamard gate on ion $m$ is defined as $U_{Hm}$: $|g
(e)\rangle_{m} \rightarrow [|g\rangle_{m} + (-) |e\rangle_{m}]/\sqrt{2}$.
All these operations can be done with high fidelity by microwave pulses
involving refocusing \cite{deng,book1}.

After a two-qubit gate CNOT$_{12}$, followed by a Hadamard gate $U_{H1}$, we
have from Eq. (6), 
\[
(1/2\sqrt{2}) |gg\rangle_{12}(|\sigma_{+}\sigma_{+}\rangle_{12} +
|\sigma_{0}\sigma_{0}\rangle_{12}) +
\]
\[
(1/2\sqrt{2})|ee\rangle_{12}(|\sigma_{0}\sigma_{+}\rangle_{12} -
|\sigma_{+}\sigma_{0}\rangle_{12}) +
\]
\[
(1/2\sqrt{2})|eg\rangle_{12}(|\sigma_{0}\sigma_{0}\rangle_{12} -
|\sigma_{+}\sigma_{+}\rangle_{12}) +
\]
\begin{equation}
(1/2\sqrt{2})|ge\rangle_{12}(|\sigma_{0}\sigma_{+}\rangle_{12} +
|\sigma_{+}\sigma_{0}\rangle_{12}).
\end{equation}
So a certain detection on the ions would yield a certain entangled photon
pair. In other words, if we have two photons produced, we can  entangle them
deterministically by our scheme.

To have a highly efficient generation of entangled photons, we require $0 <
\kappa_{m} \ll \tilde{\Omega}_{m}$. In the limit of $\kappa_{m} = 0$, the
success rate $P=1$. Current ion-trap-cavity setup could not meet our
requirement because $\kappa_{m} \gg \tilde{\Omega}_{m}$ (i.e., $%
\kappa_{m}=0.64$ MHz, $\Omega_{jm}\sim$ h$_{jm}=0.02$ MHz, and $%
\delta_{jm}=0.1$ MHz) \cite{mundt}. But in cavity QED, h$_{jm}$ and $%
\kappa_{m}$ are dependent on the size $R$ and the quality of the cavity. We
have following relations \cite{keller}: h$_{jm} \sim R^{-3/4}$ and $%
\kappa_{m} \sim (T/R)$ with T the total loss of the cavity field. For the
cavity in our case of the size 10 $\mu m$ and with the same finesses as in 
\cite{keller}, we have h$_{jm}=138.4 $ MHz and $\kappa_{m}=$ 960 MHz.
Suppose $\Omega_{jm} = 10$ MHz, and $\delta_{jm}=$ 0.1 MHz, we have $\tilde{%
\Omega}_{m}\gg \kappa_{m} $. Fig.2 presents different cases regarding
different $\delta_{jm}$ and $\kappa_{m}$.

How to collect the emitted photons is important to the application of our
scheme. The frequency difference between $|g\rangle_{m}$ and $|e\rangle_{m}$
is different from the ion to ion due to the applied magnetic field gradient.
By choosing suitable detunings, however, we can have all the generated
photons with the same frequency, but with different polarizations associated
with the deexcited levels $|g\rangle_{m}$ and $|e\rangle_{m}$. This implies
that we need two high-Q two-mode (i.e., $\sigma_{0}$ and $\sigma_{+}$)
cavities \cite{exp2} to collect the photons from the microcavities involving
the ions, respectively, as shown in Fig.1 (b). Moreover, the minimum photon
storage time is the time for the two leaking photons becoming entangled.
Table I gives $J$ of the order of KHz. So considering the time for Hadamard
gate and refocusing pulses, generation of an entangled photon pair in our
scheme takes time at least of the order of millisecond.

\section{generation of entangled multi-photon states}

Our scheme is suitable for not only repeatedly producing entangled photon
pairs, but also generating entangled multi-photon states. For example, with
three ions confined in three microtrap-microcavity setups, respectively,
after laser pulse radiations on these ions, followed by the cavity-induced
emission, we perform CNOT$_{12}$, CNOT$_{13}$, and a Hadamard gate $U_{H1}$,
which yields 
\[
(1/4)|ggg\rangle_{123}(|\sigma_{+}\sigma_{+}\sigma_{+}\rangle_{123} +
|\sigma_{0}\sigma_{0}\sigma_{0}\rangle_{123}) +
\]
\[
(1/4)|egg\rangle_{123}(|\sigma_{0}\sigma_{0}\sigma_{0}\rangle_{123} -
|\sigma_{+}\sigma_{+}\sigma_{+}\rangle_{123}) +
\]
\[
(1/4)|gee\rangle_{123}(|\sigma_{0}\sigma_{+}\sigma_{+}\rangle_{123} +
|\sigma_{+}\sigma_{0}\sigma_{0}\rangle_{123}) +
\]
\[
(1/4)|eee\rangle_{123}(|\sigma_{0}\sigma_{+}\sigma_{+}\rangle_{123} -
|\sigma_{+}\sigma_{0}\sigma_{0}\rangle_{123}) +
\]
\[
(1/4)|geg\rangle_{123}(|\sigma_{+}\sigma_{0}\sigma_{+}\rangle_{123} +
|\sigma_{0}\sigma_{+}\sigma_{0}\rangle_{123}) + 
\]
\[
(1/4)|eeg\rangle_{123}(|\sigma_{0}\sigma_{+}\sigma_{0}\rangle_{123} -
|\sigma_{+}\sigma_{0}\sigma_{+}\rangle_{123}) + 
\]
\[
(1/4)|gge\rangle_{123}(|\sigma_{0}\sigma_{0}\sigma_{+}\rangle_{123} +
|\sigma_{+}\sigma_{+}\sigma_{0}\rangle_{123}) + 
\]
\begin{equation}
(1/4)|ege\rangle_{123}(|\sigma_{0}\sigma_{0}\sigma_{+}\rangle_{123} -
|\sigma_{+}\sigma_{+}\sigma_{0}\rangle_{123}).
\end{equation}
So once we have produced the photons from the ions embedded in the
microcavities, we can generate any entangled three-photon state with a
certain probability by measuring the internal states of the ions. In
principle, for N ions confined in N microtrap-microcavity setups,
respectively, we can generate $2^{N}$ different entangled states by implementing
CNOT$_{12}$, $\cdots$, CNOT$_{1N}$, and $U_{H1}$, followed by measurements
on the internal states.

Entangling more photons, however, takes longer time because more operations
should be taken. Besides, to make our scheme work well, we hope that the
couplings between any two of the ions are on the same order of magnitude so
that we can carry out CNOT$_{ij}$ ($j > i+1$) efficiently. As shown in Table
II with different cases for three ions under consideration, the
nearest-neighbor coupling is almost equivalent to the second
nearest-neighbor one. To reduce infidelity due to the vibrations of the
ions, we restrict $\epsilon \le 0.071$ in our calculation, where $\epsilon$
is an additional Lamb-Dicke parameter regarding the magnetic field gradient 
\cite{explain1}. Therefore, the magnetic field gradient is also restricted
to be smaller than a certain number and thereby the spin-spin coupling could
not be very large. Due to the weak coupling, in addition to more steps to
take, the implementation time in three-ion case is much longer than in the
two-ion case.

More specifically, to carry out our scheme for N (e.g., $N < 10$) ions, we
need a minimum time $(N-1)\times t_{0} + t_{1}$ to generate the entangled
photons, where we suppose the time for a single CNOT$_{ij}$ to be $t_{0}$,
and other time for Hadamard gate and measurement to be $t_{1}$.
On the other hand, for more photons entangled, the success rate in our
scheme would be lower. This is due to the enhanced failure probabilities of
emitting desired photons from individual ions ($\propto$ $(1-P_{1}\cdots
P_{N})$ with $P_{m}$ the success rate for each ion), and of the detection
for a certain entangled photon states.

Although the generation rate of entangled photon pairs with our scheme
(i.e., 1 pair/ms) is lower than with other methods, e.g. in Refs. \cite%
{kwiat,bensen}, we argue that our proposal is good at generation of
entangled multi-photon states. The  previous scheme \cite{bensen} by means
of biexcitons in semiconductor quantum dots is hard to extend to creation of
entangled state for more than two photons. Although with the ideas in Refs. 
\cite{aspect,loss} entangled multi-photon states could be generated, our
scheme is advantageous to create complete sets of entangled multi-photon
states. Moverover, the proposals based on linear optical elements suffer from
low success rate because half of outputs have to be discarded. For example,
in a recent experiment for entangled five-photon states \cite{zhao},
the success rate after the photons going through several beam splitters is
lower than 1/128, even if we neglect other imperfect factors. In contrast,
with our scheme, any entangled state of five photons can be obtained in an
ideal implementation with equal success rate of 1/32.

\section{discussion and conclusion}

There are some points we have to emphasize for realizing the present scheme.
First, the long-range Coulomb force regarding the spin-spin coupling between
non-neighboring ions makes the scheme hard to scale up to a large number of
ions. We have noticed that a scalable approach is proposed in \cite{twamley}
by making the nearest-neighbor coupling prominent by properly choosing the
trapping potentials of individual traps as well as the distance between the
traps. However, in the discussion above, we prefer nearly constant coupling
strength between any two ions. So our scheme is only valid for few-photon
entanglement.

Second, although the vibrational modes of the ions remain unchanged
throughout the scheme, the spin-spin coupling we employ here is from
virtually exciting the vibrational mode \cite{twamley}, which is similar to
the idea in hot-ion scheme \cite{sm}. As an example, we estimate the
two-qubit gating time to be 2 ms, and only $10\%$ vibrational modes are
actually excited during the gating. This implies that our scheme would work
as long as the heating times of the vibrational states are longer than 0.2
ms. However, the heating rate changes with the trap size as $L^{-s}$, where
L is the trap size and s can be 2, 4 and 5 depending on different traps \cite%
{single}. Since the heating time for a trap with $L=100\mu m$ is 4 ms \cite%
{barret}, the microtrap smaller than 10 $\mu m$ would be of very short
heating time. Therefore, more advanced techniques to improve current
microtraps are highly expected to achieve our scheme.

Alternatively, we can give up the microtraps, but put the ions with each
confined in a microcavity into a linear trap. Current experiments in linear
trap have achieved the ultracold ions confined with the spacing of the
order of $\mu $m and the heating time of the order of microsec \cite{riebe}.
While in this case, the ions experience the same trap frequency, and thereby
we will not have a nearly constant spin-spin coupling between any two ions.
This might more or less prolong the implementation time, but will not be a
serious problem for the few-ion case anyway.

Third, by choosing different detunings, we may generate photons with
different energies (i.e., frequencies) but with polarization entangled, which 
are also useful \cite {bensen}. But in this
case, we need microcavities with two modes of different frequencies and
different polarizations to generate the photons and to collect the leaking
photons, which makes the experimental requirements more challenging.  

Fourth, to make our scheme work well, we should have the Lamb-Dicke
parameter of each ion much smaller than 1 to avoid any vibrational mode
excitation. As referred to above, however, besides the usually defined
Lamb-Dicke parameter $\eta $ related to the laser radiation, there is an
additional contribution $\epsilon $ in our case from the magnetic field
gradient. Provided $\eta =0.1$, the effective Lamb-Dicke parameter $\eta
_{1}=\sqrt{\eta ^{2}+\epsilon ^{2}}\leq 0.12$, where we have used the
numbers in Table 1. Anyway, a larger Lamb-Dicke parameter, even if a little
bit larger, is not good for our implementation of the scheme.

Finally, although we have mentiond in \cite{exp2} that no detailed
discussion regarding the techniques of photon collection and storage would
be given in this paper, we have to point out that both the collection and
storage of the entangled multi-photon would be more difficult than that of
photon pairs. Since the leaking photons from the mirrors of each microcavities 
would be directed toward the corresponding cavities for photon collection, we neglected 
any imperfect factors due to the devices in above discussion and assumed 
 that the collection rate is unity.  As known from \cite{keller}, 
however, the bigger the cavity, the longer
survival time the inside photons. Because the implementation time in the
multi-photon case is much longer than the case of photon pairs, we have to
have much bigger cavities for collecting leaking photons in the multi-photon
cases. This will probably reduce the collection rate due to the size
difference between the microcavities for photon generation and cavities for
photon collection. For example, the current optical cavity (of the size of 0.8 
$\sim $ 10 mm) is bigger than our required cavity for photon collection by at 
least 80 times \cite {mundt,keller}. To store the photons with high fidelity 
in the three-photon case, however, by using the numbers in Tables I and II, 
we require the size of the currently available optical cavity to be further 
enlarged by at least 5 times. This big difference in
size is evidently not good for highly efficiently collecting the leaking
photons, and the solution of this problem relies on the further advance in
cavity QED: Reducing the cavity size by two orders of magnitude and increasing
the finesses of the cavity by two to three orders of magnitude \cite {zou}.

In summary, we have proposed a potentially practical scheme to generate
entangled photons by trapped-ions embedded in microcavities in a magnetic
field gradient. We use microtraps to fix the ions, and by adjusting the
microtraps we can have different spin-spin couplings. The (micro)cavities play
the role in photon generation and collection. Compared to other previous
proposals \cite{papers}, our scheme gives a reverse step, and could provide
an efficient generation of entangled photons. Although the necessary
microtraps and microcavities are not within reach of the present techniques,
our proposal has advantages of creating complete sets of entangled
multi-photon states, which would be a stable and efficient source of
entangled photons useful for future large-scale quantum information
processing.

\section{acknowledgement}

One of the authors (MF) is grateful to D. Mc Hugh, J. Twamley and C.
Wunderlich for discussion. The work is supported by National Natural Science
Foundation of China with the contract Nos. 10474118 and 10274093, and by
National Foundamental Research Program of China Nos. 2001CB309309 and
2005CB724502.

\textbf{Note added}: After finishing this paper, we become aware of a work 
\cite{cirac}, in which the Ising coupling between ions is obtained by
changing the laser intensities and the polarizations. As it is
mathematically identical to the models in \cite%
{wunderlich1,wunderlich2,twamley}, our scheme can in principle be applied to
it.

\newpage

\begin{quote}
\textbf{Table I}. Ising coupling J in two-ion case, where $d$ is the
center-to-center distance between the two microtraps, $\nu_{m}$ $(m=1,2)$ is
the frequency of the microtrap $m$, $\Delta_{m}$ $(m=1,2)$ is the deviation
of the equilibrium position of ion 1 or ion 2 from their respective trap
centers, and $h=d+\Delta_{1}+\Delta_{2}$ is the inter-ion distance. $\epsilon
$ is the biggest one in $\epsilon_{nl}$ which is related to the nth ion and
the lth vibrational mode. The listed J is the largest spin-spin coupling in
our calculation under $\epsilon < 0.071$ and a specific $d$.

\begin{tabular}{|l|c|c|c|c|c|c|c|r|}
\hline
d $(\mu m)$ & $\nu_{1}$ (MHz) & $\nu_{2}$ (MHz) & $\partial B/\partial z
(T/m)$ & $\Delta_{1} $$(\mu m)$ & $\Delta_{2}$ $(\mu m)$ & h $(\mu m)$ & $%
\epsilon$ & J (KHz) \\ \hline
6.0 & 5.55 & 5.55 & 550 & 0.521 & 0.521 & 7.042 & 7.066e-002 & 6.328 \\ 
\hline
7.0 & 4.50 & 4.50 & 400 & 0.588 & 0.588 & 8.176 & 7.038e-002 & 4.980 \\ 
\hline
8.0 & 3.75 & 3.75 & 300 & 0.653 & 0.653 & 9.307 & 6.939e-002 & 3.959 \\ 
\hline
9.0 & 3.30 & 3.30 & 250 & 0.681 & 0.681 & 10.362 & 7.005e-002 & 3.370 \\ 
\hline
10.0 & 2.35 & 2.35 & 150 & 1.001 & 1.001 & 12.001 & 6.994e-002 & 2.875 \\ 
\hline
\end{tabular}

\textbf{Table II}. Ising couplings in three-ion case, where $d$ is the
center-to-center distance between two neighboring traps, $\nu_{m}$ $(m=1,2)$
is the frequency of the microtrap $m$, ( $\nu_{3}=\nu_{1}$ due to the
symmetry. So we omit it). $\Delta$ is the deviation of the equilibrium
position of ion 1 or ion 3 from their respective trap centers, and $%
h=d+\Delta$ is the distance between two neighboring ions. Due to symmetry,
ion 2 (i.e., the middle one) oscillates at the center of the trap confining
it, and $J_{12}=J_{23}$.  Other parameters are defined in the text.

\begin{tabular}{|l|c|c|c|c|c|c|c|r|}
\hline
d $(\mu m)$ & $\nu_{1}$ (MHz) & $\nu_{2}$ (MHz) & $\partial B/\partial z
(T/m)$ & $\Delta $$(\mu m)$ & h $(\mu m)$ & $\epsilon$ & J$_{12}$ (KHz) & J$%
_{13}$ (KHz) \\ \hline
6.0 & 2.75 & 7.75 & 240 & 2.037 & 8.037 & 6.994e-002 & 1.455 & 1.448 \\ 
\hline
7.0 & 2.55 & 7.25 & 210 & 1.922 & 8.922 & 7.048e-002 & 1.141 & 1.149 \\ 
\hline
8.0 & 2.05 & 5.80 & 150 & 2.252 & 10.25 & 6.962e-002 & 0.922 & 0.922 \\ 
\hline
9.0 & 1.45 & 4.10 & 90 & 3.186 & 12.19 & 6.810e-002 & 0.747 & 0.747 \\ \hline
10.0 & 1.20 & 3.40 & 70 & 3.688 & 13.69 & 6.996e-002 & 0.670 & 0.672 \\ 
\hline
\end{tabular}
\end{quote}

\begin{center}
\textbf{Captions of the figures}
\end{center}

Fig .1 {(a) Level scheme for a single Yb ion, where the states regarding $F=1
$ for $S_{1/2}$ and $P_{1/2}$ are split into three levels respectively, and
the lowest level (0,0) is for $F=0, S_{1/2}$. We encode qubits in $|g\rangle$
and $|e\rangle$, respectively, and take $|r\rangle$, $|g^{\prime}\rangle$,
and $|e^{\prime}\rangle$ to be auxiliary states. $\Omega_{jm}$ and h$_{jm}
(j=g, e$ and $m=1, 2)$ are coupling constants regarding lasers and cavity
modes, respectively. $\sigma_{0}$ and $\sigma_{+}$ are for polarization of
the classical and quantum fields. $\delta_{jm}$ is the detuning. (b)
Schematic plot for generation of two leaking photons from the two-mode
microcavites (on the left) involving the ions by laser radiation and for
collection by two two-mode cavities (on the right) with high-Q, where $B(z)$
is the magnetic field gradient, and the two ions are confined in two
microtraps, respectively.\newline
}

Fig.2 Success rate of two-photon generation vs. cavity dacay rate $\kappa$
based on Eq. (5) for $P=P_{1}P_{2}$, where for simplicity we suppose the
situation in the two microcavities to be the same, i.e., $%
\Omega_{j1}=\Omega_{j2}=\Omega$, h$_{j1}=$h$_{j2}=$H, and $%
\delta_{j1}=\delta_{j2}=\delta$. We set H$=138$ MHz, $\Omega=$ 10 MHz. The
curves from top to bottom correspond to $\delta=$ 0.1 MHz, 0.25 MHz, 0.5
MHz, and 1.0 MHz. If $\delta \le 10^{-3}$ MHz, P is almost 1 even when $%
\kappa=10^{9}$ Hz. For generating two photons with the same energy but
entangled in polarization, $\delta_{em}$ should be different from $%
\delta_{gm}$. While in that case the results are very similar to those shown
here.


\begin{thebibliography}{99}
\bibitem{aspect} A. Aspect, P. Grangier, and G. Roger, Phys. Rev. Lett. 
\textbf{49}, 91 (1982).

\bibitem{kwiat} P.G. Kwiat, K. Mattle, H. Weinfurter, A. Zeilinger, A.V.
Sergienko, and Y.H. Shih, Phys. Rev. Lett. \textbf{75}, 4337 (1995).

\bibitem{bensen} O. Bensen, C. Santori, M. Pelton, and Y. Yamamoto, Phys.
Rev. Lett. \textbf{84}, 2513 (2000).

\bibitem{loss} V. Cerletti, O. Gywat, and D. Loss, eprint quant-ph/0411235.

\bibitem{shih} Y.H. Shih and C.O. Alley, Phys. Rev. Lett. \textbf{61}, 2921
(1988).

\bibitem{papers} For example, M.B. Plenio \textit{et al}, Phys. Rev. A 
\textbf{59}, 2468 (1999); A.S. Sorensen and K. Molmer, Phys. Rev. Lett. 
\textbf{90}, 12703 (2003); L.-M. Duan and H.J. Kimble, Phys. Rev. Lett. 
\textbf{90}, 253601 (2003).

\bibitem{beige} Y. Liamg and A. Beige, J. Phys. A \textbf{38}, L7 (2005).

\bibitem{wunderlich1} F. Mintert and C. Wunderlich, Phys. Rev. Lett. \textbf{%
87}, 257904 (2001).

\bibitem{wunderlich2} C. Wunderlich, \textit{in Laser Physics at the limit},
edited by H. Figger, D. Mesched and C. Zimmermann (Springer Verlag, Berlin,
2001)) P261.

\bibitem{twamley} D. Mc Hugh and J. Twamley, Phys. Rev. A \textbf{71}, 12315
(2005).

\bibitem{feng} E. Biolatti, R. Lotti, P. Zanardi, and F. Rossi, Phys. Rev.
Lett. \textbf{85}, 5647 (2000); M. Feng, I. D'Amico, P. Zanardi and F.
Rossi, Europhys. Lett. \textbf{66}, 14 (2004).

\bibitem{zou} X. Zou and W. Mathis, Phys. Rev. A \textbf{71}, 042334 (2005).

\bibitem{riebe} M. Riebe, H. H\"{a}ffner, C.F. Roos, W. H\"{a}nsel, J.
Benhelm, G.P.T. Lancaster, T.W. K\"{o}rber, C. Becher, F. Schmidt-Kaler,
D.F.V. James and R. Blatt, Nature (London) \textbf{429}, 734 (2004).

\bibitem{barret} M.D. Barrett, J. Chiaverini, T. Schaetz, J. Britton, W.M.
Itano, J.D. Jost, E. Knill, C. Langer, D. Leibfried, R. Ozeri, and D.J.
Wineland, Nature (London) \textbf{429}, 737 (2004).

\bibitem{deng} Z.J. Deng, M. Feng and K.L. Gao, to appear in Phys. Lett. A;
Also available in eprint quant-ph/0507080.

\bibitem{blatt} R. Blatt and P. Zoller, J. Eur. Phys. \textbf{9}, 250 (1998).

\bibitem{book1} M.A. Nielsen and I.L. Chuang, \textit{Quantum Computation
and Quantum Information}, (Cambridge University Press, 2000).

\bibitem{mundt} A.B. Mundt, A. Kreuter, C. Becher, D. Leibfried, J. Eschner,
F. Schmidt-Kaler, R. Blatt,  Phys. Rev. Lett. \textbf{89}, 103001 (2002).

\bibitem{keller} M. Keller, B. Lange, K. Hayasaka, W. Lange and H. Walther,
New J. Phys. \textbf{6}, 95 (2004).

\bibitem{exp2} Since the photons leaking out of the two microcavities are of
the same energy (or say, frequency), but probably of different
polarizations, the two collection cavities should also be of two modes and
are designed to be resonant with the photons getting in. However, the
current optical cavity decays on the order of millisec \cite{mundt}, which is
of the same order as the minimum storage time calculated below for the
entangled photon pairs. Alternatively, we may use two high-Q fibers to
collect the photons. While according to the minimum storage time calculated
below, the fibers should be at least thousands of kilometers long.
Therefore, both of these collection methods are experimentally challenging
at present, and the detailed discussion in technique is beyond the scope of
this paper. We here argue that, even if we neglect the collection step, our
scheme is good for an efficient source of entangled photons.

\bibitem{explain1} As shown in \cite{wunderlich1,wunderlich2,twamley,deng},
this additional Lamb-Dicke parameter is related to the different confined
ions and the different vibrational modes. So it is denoted by $\epsilon_{nl}$
for the nth ion and the lth vibrational mode. Here we choose the biggest $%
\epsilon_{nl}$ to be $\epsilon$, and restrict it to be smaller than 0.071.

\bibitem{zhao} Z. Zhao, Y.A. Chen, A.N. Zhang, T. Yang, N.J. Briegel and
J.W. Pan, Nature (London) \textbf{430}, 54 (2005).

\bibitem{sm} For example, A. S$\phi$rensen and K. M$\phi$lmer, Phys. Rev.
Lett. \textbf{82}, 1971 (1999).

\bibitem{single} D. Leibfrid, R. Blatt, C. Monroe, and D. Wineland, Rev.
Mod. Phys. \textbf{75}, 281 (2003).

\bibitem{cirac} D. Porras and J.I. Cirac, Phys. Rev. Lett. \textbf{92},
207901 (2004).
\end{thebibliography}
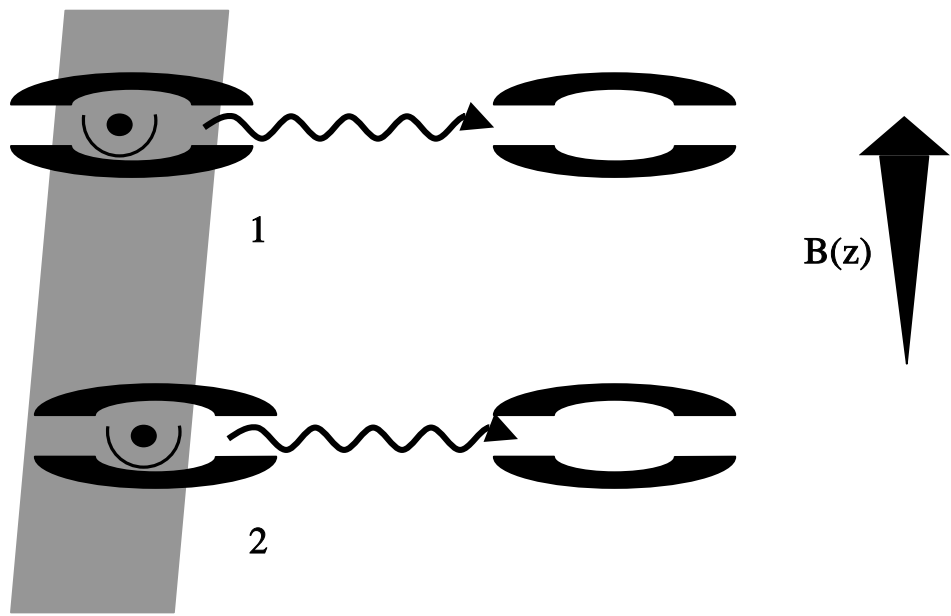
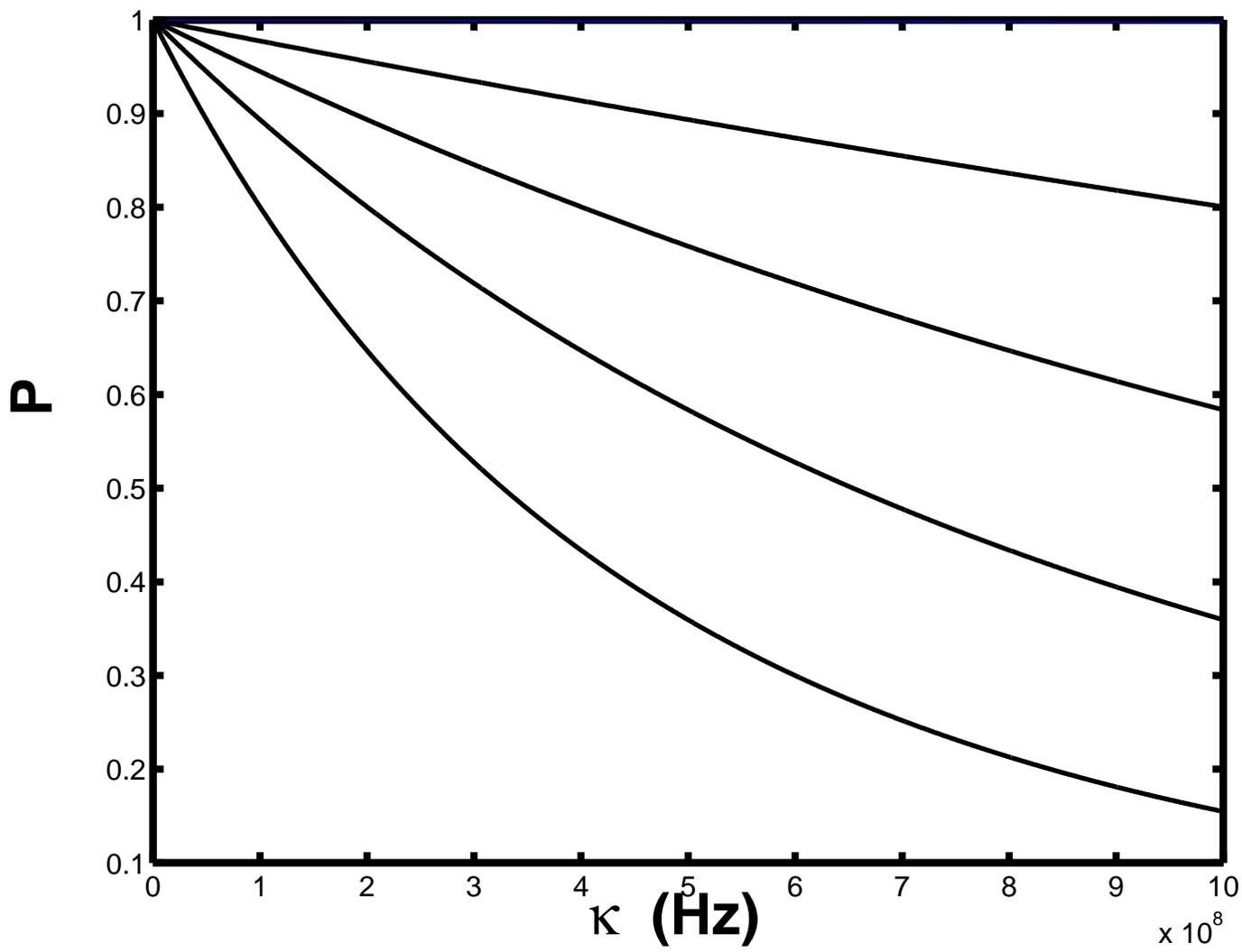
\end{document}